\documentstyle[12pt]{article}
\def\bfr{\begin{flushright}}
\def\efr{\end{flushright}}
\def\bfl{\begin{flushleft}}
\def\efl{\end{flushleft}}

\def\sii{\qquad}

\def\vs{\vspace}

\def\vsss{\vspace{1cm}}

\def\begc{\begin{center}}
\def\endc{\end{center}}
\def\bmp{\begin{minipage}}
\def\emp{\end{minipage}}

\def\bena{\begin{eqnarray}}
\def\benan{\begin{eqnarray*}}
\def\eena{\end{eqnarray}}
\def\eenan{\end{eqnarray*}}
\def\nn{\nonumber  \\}

\def\begs{\begin{screen}}
\def\ends{\end{screen}}
\def\rarw{\rightarrow}

\def\mbbl{\makebox[4cm][l]}

\def\a{\alpha}
\def\b{\beta}

\def\d{\delta}

\def\e{\epsilon}

\def\m{\mu}
\def\n{\nu}

\def\o{\o}

\def\P{\Pi}

\def\r{\rho}

\def\f{\phi}
\def\F{\Phi}
\def\vf{\varphi}
\def\c{\chi}
\def\ps{\psi}
\def\Ps{\Psi}
\def\w{\omega}

\def\fr{\frac}
\def\del{\partial}

\def\dt{\fr{\del}{\del t}}
\def\dx{\fr{\del}{\del x}}

\def\Sr{Schr\"{o}dinger}
\def\mxxb{\left( \begin{array}{cc}}           
\def\mxxe{\end{array} \right)}
\def\mxxxb{\left( \begin{array}{ccc}}         
\def\mxxxe{\end{array} \right)}
\def\mxxxxb{\left( \begin{array}{cccc}}       
\def\mxxxxe{\end{array} \right)}
\def\mxxxxxb{\left( \begin{array}{ccccc}}     
\def\mxxxxxe{\end{array} \right)}
\def\kakkob{\left\{ \begin{array}{c}}
\def\kakkoe{\end{array}
            \right. }
\def\vecb{\left( \begin{array}{c}}
\def\vece{\end{array} \right) }
\def\vectttb{\left( \begin{array}{c}}
\def\vecttte{\end{array} \right) }
\def\delp{\del _+}
\def\delm{\del _-}
\begin{document}
\bfr
\mbbl{TEP-10}\\
\mbbl{Feb. 1993}\\
\mbbl{hep-th/9302142}
\efr

 \vs{1cm}

\begin{center}
{\LARGE {\bf Wave Functional of Quantum Black Holes in Two Dimensions}}
 \vs{3.0cm}

\renewcommand{\thefootnote}{\fnsymbol{footnote}}
{\bf Takayuki Hori}\footnote[2]{E-mail address: e00353@sinet.ad.jp}
 \vs{0.5cm}

{\it Institute for Physics, Teikyo University, Otsuka 359, Hachioji-shi}
 \vs{.3cm}

{\it Tokyo 192-03, Japan } \vs{1.0cm}

and\vs{1cm}

{\bf Masaru Kamata} \vs{0.5cm}

{\it Kisarazu National College of Technology, Kiyomidai-Higashi 2-11-1,
Kisarazu-shi}\vs{.3cm}

{\it  Chiba 292, Japan} \vs{3.0cm}

{\large {\bf Abstract}}
\end{center}

\begin{normalsize}
\baselineskip=25pt
The Wheeler-DeWitt method is applied to the quantization of the 1 + 1
dimensional dilaton gravity coupled with the conformal matter fields.
Exact solutions to the WD equations are found, which are classically
interpreted as right(left)-moving black holes.
\newpage
\setcounter{page}{2}
\pagestyle{plain}
The two dimensional dilaton gravity recently attracts  much attention as
a sample model of the four dimensional theory.
Since the pioneering work by Callan et al. \cite{cghs} many authors have
studied the two dimensional models (see ref.\cite{gidd} for a review) and
attempted to unravel the mysteries concerning the black hole evaporation and
the resulting information loss \cite{hawk}, which imply a fundamental
conflict between the gravity and the quantum theories.\\
In quantizing the four dimensional Einstein gravity the canonical
procedure leads to a beautiful scheme known as the Wheeler-Dewitt (WD)
quantization \cite{whee} \cite{dwit}.

In the two dimensional dilaton gravity a WD-like equation is derived
\cite{hama} as a physical state condition.  A non-vanishing hamiltonian is
derived in a unitary gauge and a WD equation of {\Sr}  type is obtained
\cite{miko}.
There are some other approaches to the full quantum theory \cite{bila}.
As far as we know, however, an exact solution to the WD equation in the
dilaton gravity has not yet been found.

In this note we report that there exist exact functional solutions to the
WD equation of the two dimensional dilaton gravity  coupled with the
conformal matter fields (we do not consider the cosmological term).
The wave functionals are interpreted as representing quantum black holes
moving toward right or left directions, where the conformal matter fields
are present inside.

We make the assumption that the spatial dimension is compact, which seems
favourable for the cosmological applications.
The absence of the asymptotic Minkowski realm, however, brings forth a
difficult conceptual problem in interpreting the wave functional, since
then the ADM approach \cite{adm} cannot be applied.
To make contact with the extrinsic time in the WD formalism one put a time
dependent gauge condition which can be reduced to a genuine constraint by
a suitable canonical transformation.
This results in a non-vanishing hamiltonian \cite{miko}.
However, the extrinsic time has no gauge independent meaning and, in a
purely formal point of view, cannot be connected directly to physical
observables.

The interpretation of the wave functional is the outstanding problem also
in the four dimensional Einstein gravity.
We expect that the exact wave functionals, found in this note, of the
quantum dilaton gravity in two dimensions shed some light on the problem.

Now let us write the classical action \cite{cghs}
\bena
        I = \int \! d^2xe\{ e^{-2\f}[R(\w ) + 4(\nabla \f )^2]
- \fr{1}{2}\sum_{a}(\nabla f_a)^2\} ,
\eena
where $\f$ and $f_a$ are the dilaton and matter fields, respectively
(the cosmological term is omitted).
We use the zweibein formalism because it makes a calculation extremely
simple compared with the metric formalism and also is fitting for
fermionic models.
The local Lorentz gauge is fixed by putting
\bena
              e_{\m a}  =  \mxxb \a & \b e^{-\r} \\
                                    0  &    e^{\r}
                           \mxxe  ,
\eena
where $\a$ and $\b$ are the lapse function and the shift vector,
respectively.
Then the lagrangian is written as
\bena
      L = \int \! dx[e^{-2\f}\del _{\m}{\cal E}^{\m}
+ \a e^{\r}(4e^{-2\f}\del _{\m}\f \del ^{\m}\f
- \fr12\sum_{a}\del _{\m}f_a\del ^{\m}f_a)],\\
     {\cal E}^0  = -2e^{\r}K,  \sii  {\cal E}^1  =
2e^{-\r}(\b K - \a '),\\
         K  =  \fr{1}{\a}[e^{-2\r}(\b ' - \r '\b) - \dot{\r}],
\eena
where the primes and the dots denote the derivatives with respect
to $x = x^1$ and $t = x^0$, respectively.
The hamiltonian is obtained by introducing the canonical momenta
\bena
        \P _{\r} = \fr{\d L}{\d \dot{\r}},\sii
        \P _{\f} = \fr{\d L}{\d \dot{\f}},\sii
        \P _{f_a} = \fr{\d L}{\d \dot{f_a}},
\eena
then we have
\bena
         H &=& \int \! dx(\b \F _1 + \a \F _2),\\
          \F _1 &=& \e^{-2\r}[({\r}' - \dx )\P _{\r} + \f '\P _{\f}
+ \sum_{a}f'_a\P _{f_a}],\\
          \F _2 &=& e^{-\r + 2\f}[\fr14 (\P _{\r} + \P _{\f})\P _{\r}
+ \fr12e^{-2\f}\sum_{a}\P ^2_{f_a}  +  e^{-4\f}F],\\
    F &=& 4({\f '}^2 - \f ^{(2)}) + \fr{1}{2}e^{2\f}\sum_{a}f'^2_a,
\eena
where we have defined the covariant derivatives, $\f ^{(n)}$, in the
spatial dimension
\bena
            \f ^{(0)}   \equiv  \f , \sii  \f ^{(n+1)} \equiv  (\dx
- n\r ')\f ^{(n)}.
\eena

The preservation of the primary constraints $\P _{\a} = \P _{\b} = 0$
requires the secondary constraints $\F _1 = \F _2 = 0$, which form a
closed algebra \cite{miko} \cite{bila}.\\

The WD equations are obtained by the replacement, $\P _{\vf} \rarw
-i\fr{\d}{\d \vf}$ for $\vf =(\r , \f , f_a)$:
\bena
           \c \Ps [\r , \f , f_a] = 0,   \\
           \F \Ps [\r , \f , f_a] = 0,
\eena
with
\bena
        \c  &\equiv& (\r '(x) - \dx )\fr{\d}{\d \r (x)}
+ \f '(x)\fr{\d}{\d \f (x)} + \sum_{a}f_a'(x)\fr{\d}{\d f_a(x)},\\
        \F  &\equiv& \fr14 \fr{\d ^2}{\d \r (x)^2}
+ \fr14\fr{\d ^2}{\d \r (x)\d \f (x)} +
                \fr12e^{-2\f (x)}\sum_{a}\fr{\d ^2}{\d f_a(x)^2}
- e^{-4\f (x)}F(\r , \f , f_a),
\eena
where $x$ is the spatial coordinate.
The time coordinate, $t$, is implicit in the above equations, and for
a fixed $x$ the couple of the WD equations is one of the copies of an
infinite number of the same equations with various $t$.

In the present choice of variables there are apparently no ambiguities
of ordering operators, but they would exist in other choices.
This comes from the functional derivations at the same point, which should
be defined by a suitable regularization.
After all this problem is bypassed by setting $\fr{\d \vf (x)}{\d \vf (x)}
= \d (x,x) = 0$ \cite{dwit}, and we assume it in what follows for
circumventing inessential complications.
Also we drop an anomalous term in the basic WD equation, which may
possibly be appeared in other approaches \cite{hama}.\\
The $\c$-constraint, eq.(12), amounts to the spatial reparametrization
invariance of the wave functional, and its general solution is an arbitrary
function of the following functional,

\bena
       {\cal F}[\r , \f , f_a] &=& \int \! dxe^{(1 - n_1 - 2n_2 - 3n_3
- ... - m_1 - 2m_2 - 3m_3 - ...)\r }\nn
        &\times&{\f ^{(1)}}^{n_1}{\f ^{(2)}}^{n_2}{\f
^{(3)}}^{n_3}...{f^{(1)}}^{m_1}{f^{(2)}}^{m_2}{f^{(3)}}^{m_3}...\nn
        &\times&{\cal A}(\f ){\cal B}(f),
\eena
with arbitrary functions, ${\cal A}, {\cal B}$, and with arbitrary real
numbers, $n, m$'s.
While the solution to the hamiltonian constraint, eq.(13), is non-trivial,
but we found exact solutions:
\bena
          \Ps _{\pm}  =  \exp{ \int \! dx\{ e^{-2\f}[\mp 2i\f
'^{-1}\f ^{(2)}  +  M(f_1' \mp  if_2')]   +   \fr12\sum_{a = 1}^{2}\e
^{ab}f_a{f'}_b\} },
\eena
where $M$ is an arbitrary complex number.
Here we assume there are two matter fields, but it is straightforward to
extend the solution to the models with an even number of matter fields.
$\Ps _{\pm}$ satisfy also the $\c$-constraint, since ${\cal W}_{\pm}
\equiv \fr{1}{i}\log{\Ps _{\pm}}$ are of the form (16).

In order to clarify the role of the wave functionals, eq.(17), we inquire
what is the classical counterparts of them.
For that purpose let us first consider the canonical equations of motion
in the classical system.
We choose $\a = e^{\r}, \b  = 0$, corresponding to the conformal gauge
\bena
       g_{\m \n}  =  e_{\m a}{e_{\n}}^a  = e^{2\r}\mxxb -1 &  \\
                                                           & 1
                                                  \mxxe  .
\eena
Then the canonical equations of motion are
\def\piro{\P _{\r}}
\def\pifi{\P _{\f}}
\def\piff{\P _{f_a}}
\def\dpiro{\dot{\P}_{\r}}
\def\dpifi{\dot{\P}_{\f}}
\def\dpiff{\dot{\P}_{f_a}}
\bena
    \dot{\r} &=&  \fr{1}{4}e^{2\f}(2\piro  +  \pifi ),\\
    \dot{\f} &=&  \fr{1}{4}e^{2\f}\piro ,\\
    \dot{f_a} &=&  \piff ,\\
    \dpiro &=& -2(e^{-2\f})'',\\
    \dpifi  &=&  -\fr{1}{2}e^{2\f}(\piro  +  \pifi )\piro
- e^{-2\f}[8(\f '' - \f '^2) - 4\r ''],\\
    \dpiff  &=& f_a''.
\eena
Eliminating $\P$'s and using the $\c$- and the hamiltonian constraints we
get
\def\delpm{\del _{\pm}}
\bena
        \delp \delm f_a  &=&  0, \\
        \delp \delm e^{-2\f}  &=&  0, \\
       \delp \delm (\r - \f )  &=&  0,\\
       e^{-2\f}[8\delpm \r \delpm \f  - 4\delpm ^2\f ]  &+&  \sum_{a}(\delpm
f_a)^2  = 0,
\eena
where we have introduced the light-cone coordinates, $x^{\pm} = x^0 \pm
x^1$.
Since $e^{-2\f}$ is a free field we put $e^{-2\f} = u_+(x^+) + u_-(x^-)$,
and $\r  = \f  + \fr12 (w_+(x^+) + w_-(x^-))$ by eq.(27).
Then we have
\bena
        \delpm ^2u_{\pm}  -  \delpm w_{\pm}\delpm u_{\pm}
+ \fr12 \sum_{a}(\delpm f_a)^2  =  0.
\eena
Thus we can get the general solution by solving the liner equations (29)
once the matter fields, $f_a$, are given as sums of the left and the right
moving waves.

Now let us back to the wave functional.
Since we have no concept of an extrinsic time we cannot deduce the classical
{\it trajectory} directly from the exact wave functional.
Note the wave function of the non-relativistic quantum mechanics in the WKB
approximation is $\ps  = e^{iS}$ where $S$ is a solution to the
Hamilton-Jacobi equation.
Conversely, if one knows an exact solution to the {\Sr} equation one can
obtain the solution to the Hamilton-Jacobi equation simply by $S =
\fr{1}{i}\log{\ps}$, by which the classical trajectory is deduced.
Let us try in our case to deduce the {\it classical equations of motion}
by the similar procedure.

It is understood by its construction that ${\cal W} \equiv
\fr{1}{i}\log{\Ps _-}$ was determined in such a way that the {\it momenta}
defined by
\bena
        \tilde{\P}_{\r}  \equiv  \fr{\d {\cal W}}{\d \r}, \sii
\tilde{\P}_{\f}  \equiv  \fr{\d {\cal W}}{\d \f}, \sii
\tilde{\P}_{f_a}  \equiv  \fr{\d {\cal W}}{\d f_a},
\eena
satisfy the $\c$- and the hamiltonian constraints.
This implies that the canonical equations of motion with $\P$'s replaced
by $\tilde{\P}$'s become a consistent set of equations, since the
hamiltonian constraint can be regarded as the Hamilton-Jacobi equation
of ${\cal W}$.
Let us interpret the solution to the set of equations thus obtained as
the classical counterpart of the quantum wave functional.

The explicit forms of the {\it momenta} are
\bena
       \tilde{\P}_{\r}  &=&  2(e^{-2\f})',\\
       \tilde{\P}_{\f}  &=&  e^{-2\f}[-4{\f '}^{-1}(\f ^{(2)} - 2{\f '}^2)
+ 2iM(f_1' + if_2')],\\
       \tilde{\P}_{f_1}  &=&  iM(e^{-2\f})' - if_2',\\
       \tilde{\P}_{f_2}  &=&  -M(e^{-2\f})' + if_1'.
\eena
The time derivatives of $\r , \f , f$ are assumed to be {\it defined} by
eqs.(19)-(21), where $\P$'s are replaced by $\tilde{\P}$'s.
Then we have
\bena
           (\dt  +  \dx )e^{-2\f} = 0,\\
          \dot{\r} - \r '  + \fr{\f ''}{\f '}  + \fr{\mid \! \! M\! \!
\mid ^2}{2}(e^{-2\f})' = 0,\\
          \dot{f_1} = -f_1' = -(\Im M)(e^{-2\f})',\\
          \dot{f_2} = -f_2' = -(\Re M)(e^{-2\f})'.
\eena
Thus $e^{-2\f}$ and $f_a$ are right moving free fields, and we write
$e^{-2\f} = u_-(x^-)$.
The time development of $\tilde{\P}$'s must be determined by
eqs.(22)--(24), where $\P$'s are replaced by $\tilde{\P}$'s.
Hence by eq.(27) and (36)--(38) we get
\bena
        \delm ^2u_-  -  \delm w_-\delm u_-  + \fr{\mid \! \! M\! \!
\mid ^2}{2}(\delm u_-)^2 =  0.
\eena
where we have written $\r  =  \f  +  \fr12(w_+(x^+) + w_-(x^-))$.
This is a Riccati equation for $\delm u_-$ and the general solution is
\bena
             u_- = const. + \int dx^-e^{W_-}(const. + \fr{\mid \! \! M\!
\! \mid ^2}{2}\int dx^-e^{W_-})^{-1}.
\eena
The solutions of the matter fields are
\bena
              f_1 &=& f_{1-}(x^-) \equiv   (\Im M)u_-(x^-)  +  const.,\\
              f_2 &=& f_{2-}(x^-) \equiv   (\Re M)u_-(x^-)  +  const.
\eena
The striking difference between the above solutions and  the classical
ones is that the conformal matter is determined after one solves the
non-linear Riccati equation to $u_-$ in the former case, while in the
latter case the matter is first given independently then one solves the
linear equation for $u_-$.

If one choose the gauge $w_+ = w_- = 0$ then
\bena
         u_- = a  +  \fr{2}{\mid \! \! M\! \! \mid ^2}\log{\mid x^-
-  b\mid},
\eena
where $a, b$ are arbitrary constants.
This represents a right-moving black hole with the horizon at $x^- = b
\pm  e^{-\fr{a\mid M\mid ^2}{2}}$ and the matter fields are present inside.
Inside the black hole the dilaton becomes complex valued, but this simply
means that $e^{-2\f}$ becomes negative and the above formulas are valid
with no modifications.

In the same way the classical counterpart of the wave functional $\Ps _+$
turns out to a left-moving black hole.
Hence the superposition $c_1\Ps _+ + c_2\Ps _-$ is interpreted as
representing a universe in which right- and/or left- moving black holes
are observed in a certain opportunity determined by the coefficients,
$c_1$ and $c_2$.\vs{1cm}
\\
We thank H. Kakuhata and T. Shimizu for a collaboration in an early
stage of the work.
We are grateful to the members of the Doy\= o-Kai for helpful discussions.
We also thank the High Energy Group of Tokyo Metropolitan University for
hospitality during our visit.
\newpage
 \vsss

\newpage
\end{normalsize}
\end{document}